\documentclass[sort&compress]{elsarticle}

\usepackage{amsmath}
\usepackage{amssymb}
\usepackage{graphicx}
\usepackage{latexsym}
\usepackage{listings}
\usepackage{tabularx}
\usepackage{dcolumn}

\usepackage[squaren]{SIunits}
\usepackage{binary}

\usepackage{aas_macros}

\newcommand{\mvect}[1]{\mathbf{#1}}

\addunit{\FLOP}{FLOP}

\journal{Computer Physics Communications}

\begin{document}


\begin{frontmatter}

\title{Enabling a High Throughput Real Time Data Pipeline for a Large Radio Telescope Array with GPUs}

\author[iic]{R. G. Edgar}
\ead{rge21@seas.harvard.edu}

\author[iic,cfa]{M. A. Clark}
\ead{mikec@seas.harvard.edu}

\author[seas]{K. Dale}
\ead{dale@eecs.harvard.edu}

\author[cfa]{D. A. Mitchell}
\ead{dmitchell@cfa.harvard.edu}

\author[cfa]{S. M. Ord}
\ead{sord@cfa.harvard.edu}

\author[cfa]{R. B. Wayth}
\ead{rwayth@cfa.harvard.edu}

\author[seas]{H. Pfister}
\ead{pfister@seas.harvard.edu}

\author[cfa]{L. J. Greenhill}
\ead{greenhill@cfa.harvard.edu}

\address[iic]{Initiative in Innovative Computing, 29 Oxford Street, Cambridge MA 02138}
\address[cfa]{Harvard-Smithsonian Center for Astrophysics, 60 Oxford Street MS-42, Cambridge, MA 02138}
\address[seas]{School of Engineering \& Applied Sciences, 33 Oxford Street, Cambridge, MA 02138}

\begin{abstract}
The Murchison Widefield Array (MWA) is a next-generation radio telescope currently under construction in the remote Western Australia Outback.
Raw data will be generated continuously at \unit{5}{\gibi\byte\usk\reciprocal\second}, grouped into \unit{8}{\second} cadences.
This high throughput motivates the development of on-site, real time processing and reduction in preference to archiving, transport and off-line processing.
Each batch of \unit{8}{\second} data must be completely reduced before the next batch arrives.
Maintaining real time operation will require a sustained performance of around \unit{2.5}{\tera\FLOP\usk\reciprocal\second} (including convolutions, FFTs, interpolations and matrix multiplications).
We describe a scalable heterogeneous computing pipeline implementation, exploiting both the high computing density and FLOP-per-Watt ratio of modern GPUs.
The architecture is highly parallel within and across nodes, with all major processing elements performed by GPUs.
Necessary scatter-gather operations along the pipeline are loosely synchronized between the nodes hosting the GPUs.
The MWA will be a frontier scientific instrument and a pathfinder for planned peta- and exascale facilities.

\end{abstract}


\end{frontmatter}


\section{Introduction}

Many sciences are increasingly subject to a ``data deluge'' \cite{DataDeluge} which must be channeled into meaningful scientific results.
In the field of astronomy, the advent of synoptic survey and other telescopes with very wide fields of view is creating such a phenomenon.
Exacerbating the increasing data volume is the frequent need to site the telescopes far from interfering sources (e.g. radio transmitters, city lights), on isolated mountain tops, in deserts and even Antarctica.
Such locations are typically poorly served by high speed fibre links, motivating the construction of data analysis engines close to the acquisition hardware.
The Murchison Widefield Array (MWA) in an excellent example of this trend.

\begin{figure*}
\centering
\begin{tabular}{cc}
\includegraphics[height=5cm]{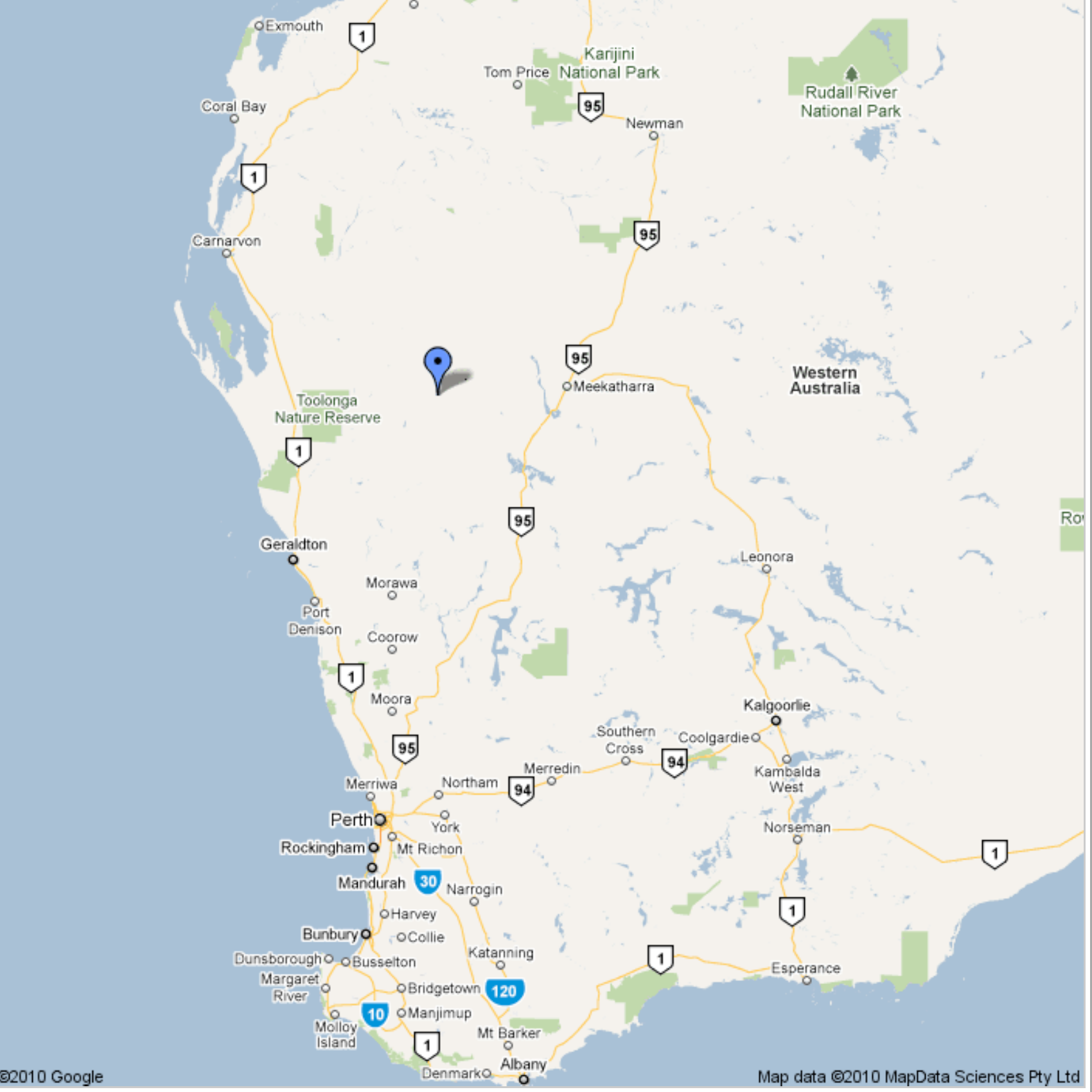} & \includegraphics[height=5cm]{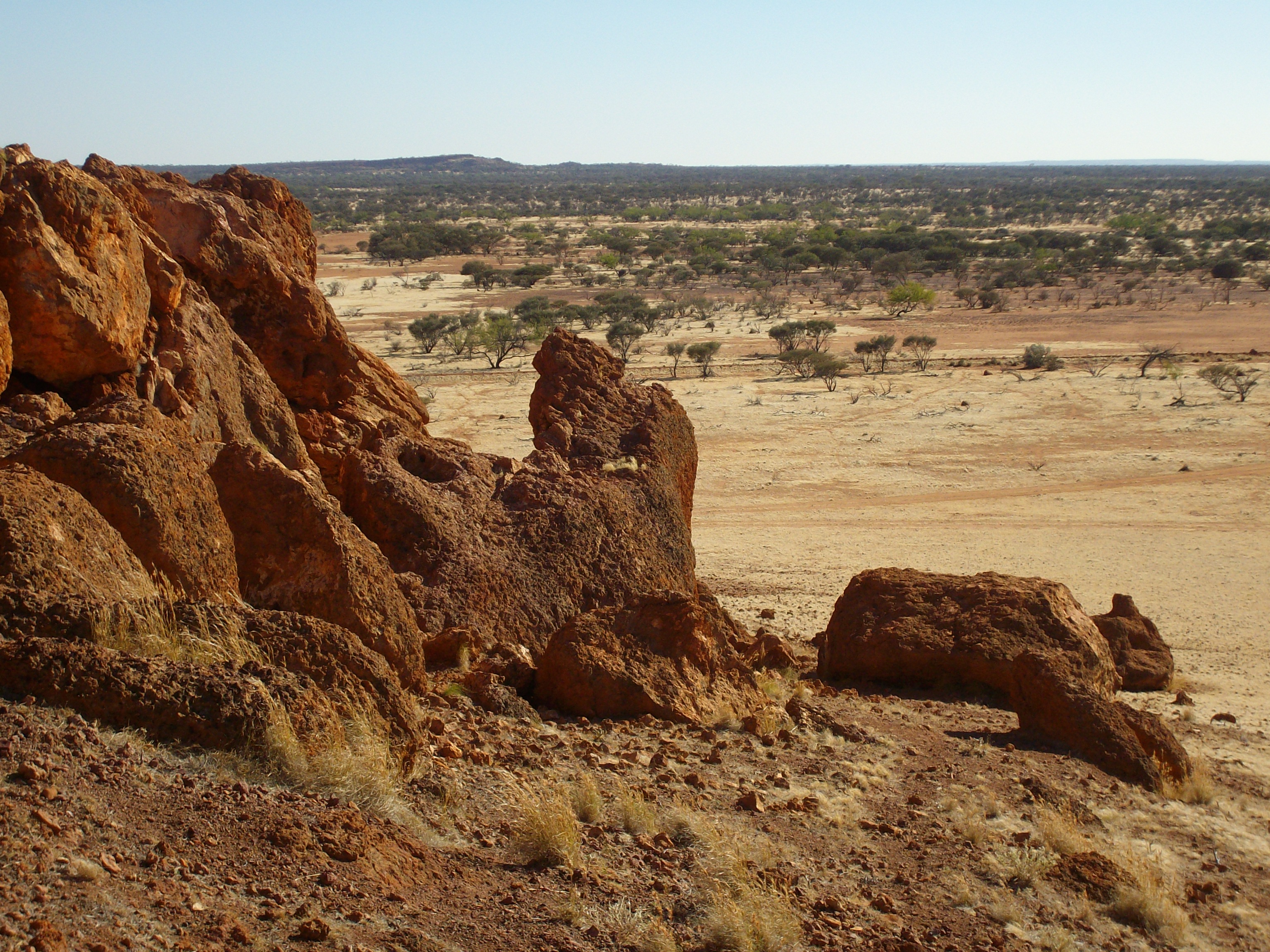}
\end{tabular}
\caption{Left: The Murchison Widefield Array is located close to Boolardy Station in Western Australia (map courtesy of Google). Right: Photograph of the telescope site.}
\label{fig:MWAlocation}
\end{figure*}

The MWA is a next-generation radio telescope being built near Boolardy Station in the Western Australian Outback (Figure~\ref{fig:MWAlocation}).
It will be used to study the early universe, the Sun, space weather, and transient radio phenomena\cite{2009IEEEP..97.1497L}.
The telescope will operate in the 80 to \unit{300}{\mega\hertz} waveband.
Cosmologists will use the MWA to map matter in the Universe during the Epoch of Reionization soon after the Big Bang, when the earliest stars, galaxies, and quasars formed \cite{2006PhR...433..181F,2009arXiv0910.3010M}.
That is the MWA's job by night, when the environment is most radio quiet.
The MWA's job by day will be mapping the Sun, with particular focus on solar storms and the ever-changing solar magnetic field that shapes those storms and in turn determines space weather around Earth \cite{2006AGUSMSH33C..03S}.
The input data stream will be \unit{5}{\gibi\byte\usk\reciprocal\second} (around \unit{3}{\pebi\byte} per week), which is too large to be streamed elsewhere for processing without an expensive, dedicated fibre link.
Instead the data must be reduced on site, which is a computationally expensive task.
We estimate that a sustained computation rate of \unit{2.5}{\tera\FLOP\usk\reciprocal\second} will be required to run the data processing pipeline in its desired configuration.

An arid, isolated location is somewhat unconventional for a supercomputer, and it presents a number of unique challenges for the MWA.
Underlying most of these is supply of electrical power.
Far from the power grid, electricity will be supplied by on-site diesel generators.
A maximum of \unit{40}{\kilo\watt} is available for the on-site data centre (exclusive of cooling), but keeping power consumption below this is strongly desirable.

Graphical Processing Units (GPUs) have evolved into highly parallel compute engines, capable of performing general purpose computations.
We leverage the high computational power (and low power consumption per \FLOP) of GPUs for the highly parallel mathematical operations of the data pipeline, while applying the strengths of CPUs to data flow control and supporting specialised calculations (e.g. ephemerides).
In this paper, we present the results of our \emph{initial} port of the MWA's data processing pipeline to NVIDIA GPUs, using the CUDA environment in a relatively straightforward fashion.
We have implemented the entire pipeline in a basic form on the GPU, since the intrinsic bandwidth demands are sufficiently high that minimizing transfers between the CPU and GPU is essential.

\section{The Murchison Widefield Array}
\label{sec:mwa}

\begin{figure}
\centering
\includegraphics[scale=0.3]{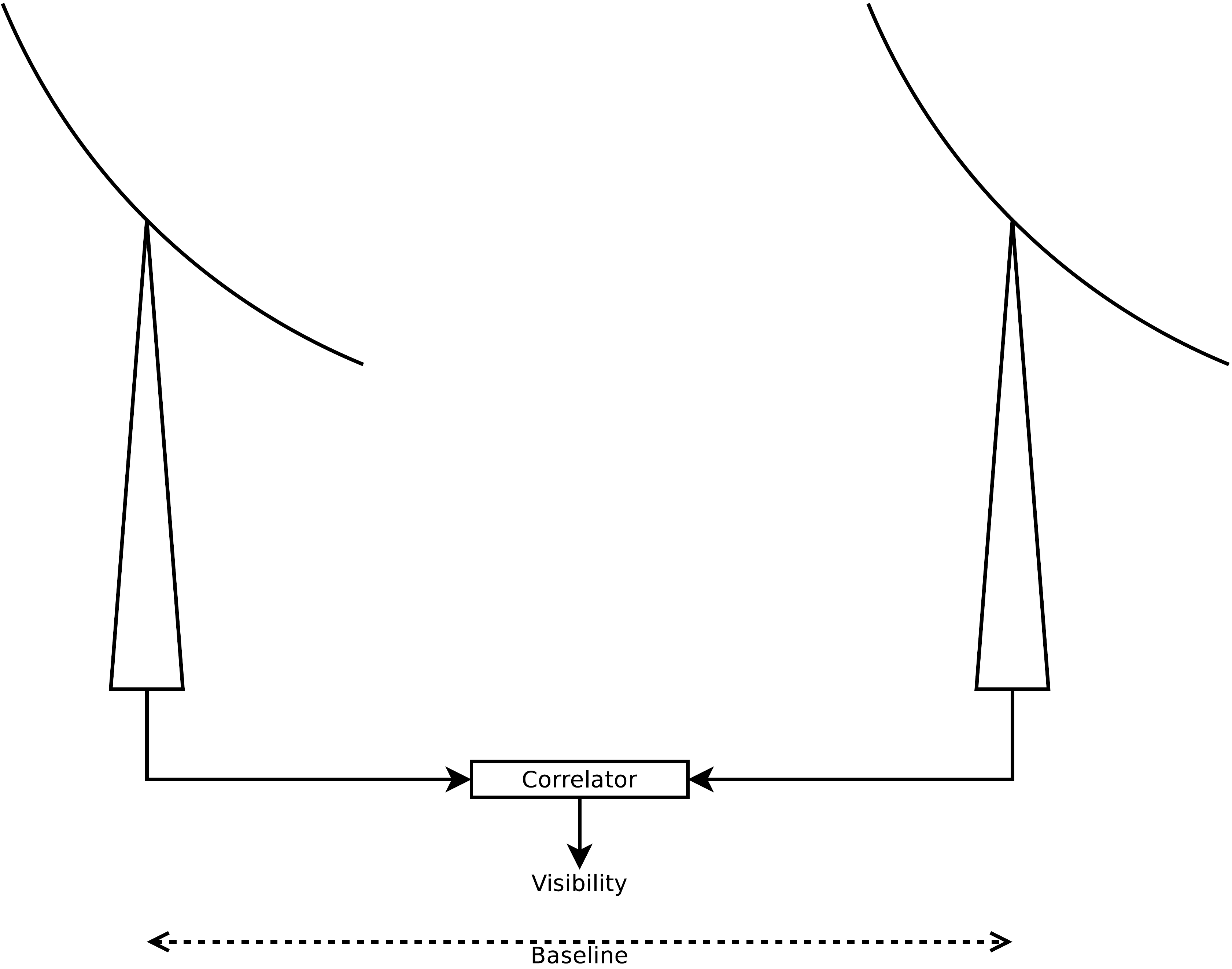}
\caption{A simple two receptor interferometer. Signals from the two receptors are combined in the correlator to produce a visibility, a component of the Fourier transform of the sky. The vector on the ground connecting the two receptors is called a baseline.}
\label{fig:simpleinterferometer}
\end{figure}

The MWA is a radio interferometer, performing simultaneous observations in 768 frequency ``channels'' (each with a \unit{40}{\kilo\hertz} bandwidth) distributed between 80 and \unit{300}{\mega\hertz}.
Measurements from multiple receptors distributed over $\sim \unit{1}{\kilo\metre\squared}$ are cross-correlated, providing a Fourier representation of the spatial distribution of flux across the radio sky.
In a snapshot, one obtains $O(n^2)$ samples of the Fourier plane for $n$ receptors (Figure~\ref{fig:simpleinterferometer}).
The samples are known as ``visibilities,'' and the Fourier plane is known as the ``$u$--$v$ plane.''
In general, more samples in the $u$--$v$ plane result in better point-source response and in turn, better image quality.
This technique is known as synthesis imaging, and a good overview of the subject is given in \cite{2001isra.book.....T}.

For the MWA, each receptor comprises a $4\times4$ grid of crossed dipole antenn\ae{} that are sensitive to two linear polarizations\cite{2009AAS...21347407C}.
The field of view of each dipole covers the entire sky.
Each grid is referred to as a ``tile.''
Within each, time delays are applied to the outputs of the dipoles, enabling them to be phased up to operate as a single \emph{aperture} with a collecting area over an order of magnitude greater than a single dipole and an electronically steerable field of view (typically around \unit{20}{\degree} across).
The MWA is distinctive in this respect, as most radio interferometers rely on steerable reflectors.
Each tile provides a single input to the correlator for each of the 768 channels.
The correlator combines pairs of tiles to make ``baselines'' and forms the visibility corresponding to each.
These broad operating parameters of the MWA are summarised in Table~\ref{tbl:mwavitalstat}.

\begin{table}
\begin{tabular}{ll}
Tiles & 512 \\
Baselines & 130816 \\
Channels & 768 \\
Channel Bandwidth & \unit{40}{\kilo\hertz} \\
Frequencies Covered & 80 to \unit{300}{\mega\hertz} \\
Field of view & $\approx \unit{20}{\degree}$ (frequency dependent) \\
Data Rate & \unit{5}{\gibi\byte\usk\reciprocal\second} \\
Operating Cadence & \unit{8}{\second}
\end{tabular}
\caption{Vital statistics of the Murchison Widefield Array}
\label{tbl:mwavitalstat}
\end{table}

The MWA has an order of magnitude more correlator inputs than other present-day instruments, such as the Very Large Array in New Mexico \cite{1980ApJS...44..151T,1981tesc.book....1H,2009AAS...21347405M} and Allen Telescope Array in California \cite{2004SPIE.5489.1021D,2004ExA....17...19D}, and a  field of view which is also an order of magnitude greater.
The computational burden in the collective calibration and imaging task includes terms that scale quadratically with the number of receptors or the field of view, meaning that the MWA data processing pipeline presents an extreme computational challenge.
We describe the pipeline system responsible for these calculations in the next section.
It functions with a cadence of \unit{8}{\second}, which is the turnover timescale for the ionospheric eddies relevant to the MWA's waveband.
These eddies introduce sky-position dependent refraction, producing distortion akin to that caused by ripples on the surface of a swimming pool.

Real time processing is a significant advance for synthesis imaging in radio astronomy.
Interferometers with tens of receptors typically rely on real time correlation and off-line computation to generate calibration solutions and sky images via iterative Fourier techniques.
Snapshot image synthesis was proposed by  \citet{1984iimp.conf..177B} as a solution to the problem of synthesising wide field images (see also \citet{EVLAmemo67}), but the computational demands rendered it impractical.
Traditionally, the processing of radio astronomy data has been a manually intensive and time consuming operation.
The visibility data taken from the telescope cannot simply be Fourier transformed into an image.
The instrument must also be calibrated and the ionospheric distortion removed.
Even the Fourier transform must take account of the incomplete sampling in the $u$--$v$ plane.
The \textsc{clean} algorithm described by \citet{1974A&AS...15..417H} serves as the basis of contemporary radio interferometry.
Although highly successful, it is an iterative and interactive approach, unsuitable for use by the MWA.
Real time, fully automatic processing is a key requirement for the instrument, meaning that the entire pipeline must use a deterministic, single pass algorithm.
The MWA required the development of new algorithms, implemented on novel hardware.

\section{The Real Time System}
\label{sec:pipeline}

The MWA's Real Time System (RTS) is responsible for processing the visibility data to finished images through a multi-stage pipeline\cite{2007AAS...211.1103M}, which consists of the following primary stages:
\begin{description}
\item[Visibility Integrator]Ingests \unit{0.5}{\second} visibility data from the correlator and integrates it into 2, 4 and \unit{8}{\second} groupings. Visibilities corresponding to widely separated tiles are integrated for shorter times. This avoids coherence loss due to the more rapid variation in time-of-arrival delays induced by the rotation of the Earth. A visibility integrated for two seconds will appear four times in a given \unit{8}{\second} cadence, each in a slightly different location in the $u$--$v$ plane.
\item[Calibration-Measurement Loop (CML)]Measures the gain of each tile and the ionospheric distortion using catalogued point sources.
The calibration problem is reduced to a set of linear equations for each calibration source.
Parameters are position offsets due to ionospheric refraction and $2\times2$ Jones matrices for the polarised response of each tile in the direction of the calibrator.
The CML steps sequentially through a list of calibrator sources, constraining a direction-dependent gain model of each tile according to an algorithm described in \cite{2008ISTSP...2..707M}.
\item[Gridder]Arranges the visibilities, scattered across the $u$--$v$ plane onto a regular grid. This is achieved by convolving each visibility with a carefully chosen gridding kernel. Ultimately, this kernel will have to be specific to each visibility, since it is determined by the calibration of the pair of tiles which contribute to each visibility.
\item[Imager]Converts the visibilities (in Fourier space) into images using an FFT.
\item[Polarisation Convertor]Removes the direction-dependent response of the dipole antenn\ae{}.
The task is split into precomputation and application steps.
The application step performs the operation $\mvect{S}_i = A_i \mvect{P}_i$ for each pixel, where $\mvect{P}_i$ is a vector containing the four observed polarisations for pixel $i$, $\mvect{S}_i$ is the deprojected vector, and $A_i$ is a $4\times4$ complex matrix representation of the transform between the two vectors.
The precomputation step supplies $A_i$ for each pixel.
\item[Regridder]Resamples the image from an instantaneous flat sky projection to the curved celestial sphere, and corrects the ionospheric distortion.
This stage is also split into precomputation and application steps.
The precomputation defines a many-to-many mapping between pixels, which is then used by the application step to redistribute the measured radio fluxes.
\end{description}
Table~\ref{tbl:datasize} shows some of the data structure sizes which are used in the pipeline.
Most figures are approximate, with the exact values depending on the precise parameters of an observing run.
Note that the size of the ionospheric fit data (which must be shared between nodes, see further discussion below) is very small.

\begin{table}
\begin{tabular}{ll}
Visibility Data   & \unit{6.5}{\mebi\byte} per channel \\
Image Data & \unit{20}{\mebi\byte} per channel\\
Calibration Data & \unit{5}{\mebi\byte} per node\\
Ionospheric Data shared for ionospheric fit & \unit{300}{\byte} per node \\
Polarisation Conversion Data & \unit{162}{\mebi\byte} per node \\
Regridding Data  & \unit{12}{\mebi\byte} per node
\end{tabular}
\caption{Approximate data sizes for the different stages of the MWA pipeline. All quantities are per \unit{8}{\second} cadence. We anticipate that each node
will process twelve channels of data}
\label{tbl:datasize}
\end{table}

\begin{figure}
\centering
\includegraphics[height=10cm]{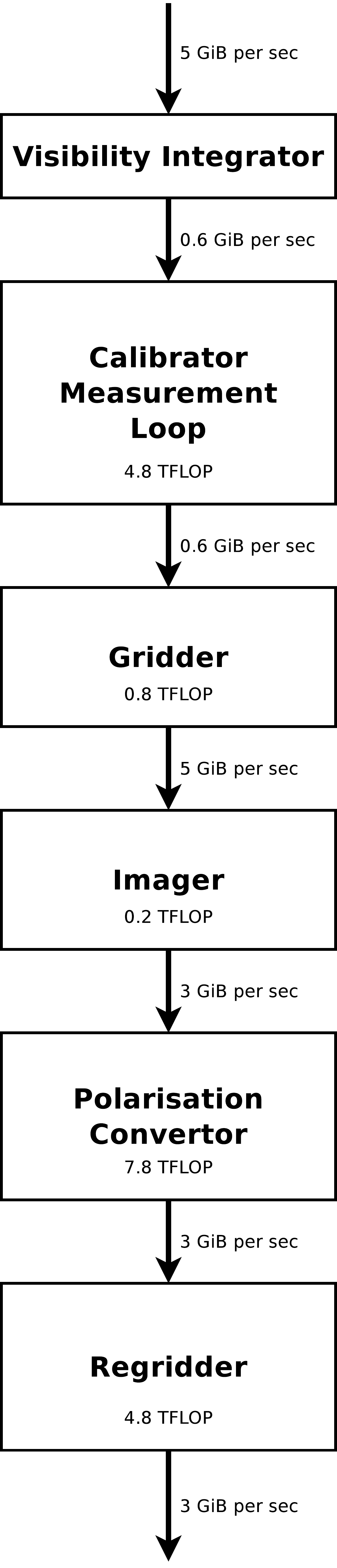}
\caption{Schematic flow chart of the Real Time System, showing the main pipeline stages.
Estimates based on 50 calibration sources within the CML, an $11\times11$ gridding kernel, producing $1125\times1125$ pixel final images and using a flux-conserving regridding method.
We do not provide a \FLOP{} estimate for the visibility integrator, since it is dominated by the cost of networking with the correlator.
The pipeline must be completed within \unit{8}{\second}}
\label{fig:RTSflow}
\end{figure}

Figure~\ref{fig:RTSflow} is a flow chart of the RTS, including estimated data rates and \FLOP{} requirements for each pipeline stage (there is no estimate for the visibility integrator, since most of the work in that stage is in networking with the correlator).
The computational budget is based on strenuous but reasonable operating parameters and algorithms, and a complete set of data must flow through the pipeline every \unit{8}{\second}.
Dependent on the parameters selected, the exact \FLOP{} budget is uncertain to within a factor of two.
Although comparing theoretical \FLOP{} needs to real processors is fraught with difficulties, it is evident that the MWA will require an on site supercomputer capable of sustaining around \unit{2.5}{\tera\FLOP\usk\reciprocal\second} in order to ensure success.

The data rate of the final images is still approximately \unit{3}{\gibi\byte\usk\reciprocal\second}, comparable to the visibility data rate from the correlator.
However, the processed images can be averaged together, reducing the final data volume.
This is not possible with the raw visibility data.
The number of \unit{8}{\second} cadences which must be averaged will be determined by the on site storage capacity and frequency of site visits.
It is expected that a `snapshot' mode will be provided where \unit{8}{\second} data is saved, to allow high time resolution analysis of transient events.

The RTS is a very parallel system.
Each of the 768 channels is almost entirely independent of the others.
The exception is within the CML, where both the calibration and ionospheric measurements benefit from combining data from multiple channels.
Within each channel there is finer grained parallelism, operating at the level of individual visibilities or pixels but with frequent synchronisations required.
We make use of this split in our implementation of the RTS and the design of the Real Time Computer.

\subsection{The Real Time Computer}
\label{sec:RTCdesc}

The Real Time Computer (RTC) will be responsible for running the RTS, and its design is heavily influenced by the unique needs of the RTS.
In the baseline configuration, there will be 32 worker nodes, each accepting two 1000BASE-T cables from the FPGA correlator.
Each worker node ingests data for a fixed set of 24 channels from the correlator, and runs it through the RTS pipeline.
At this level, we are pursuing weak parallel scaling; adding more nodes (up to a maximum of 32) enables more channels to be processed.
Within each worker node are two GPUs, which will each perform the pipeline calculations for 12 channels.
Here we require strong scaling, since we have a fixed amount of work to be completed in the smallest possible time (and certainly within \unit{8}{\second}).

The channels given to each GPU will be consecutive in frequency.
This choice is driven by the requirements of the CML.
The calibration benefits from combining consecutive channels (since the unknown tile gain may then be Taylor-expanded in frequency), while the ionospheric measurements work best with widely spaced frequencies (since the response of the ionosphere is a known function of frequency).
The ionospheric fit requires six floating point numbers per source (the fit itself is described by two floating point numbers per source), while the visibility data associated with a single channel is over \unit{6}{\mebi\byte} (see Table~\ref{tbl:datasize}).
It is therefore best to perform the calibration within a single GPU, and share the ionospheric data between GPUs.

Our early tests suggest that a full core must be dedicated to managing the UDP packets streaming from the correlator along each input cable.
We anticipate that each node will contain a single hexa-core processor, to ensure that the data flow can be managed smoothly.

The master node is responsible for constructing the ionospheric solution from data supplied by the workers, and supplying the polarisation conversion and regridding precomputations.
The ionospheric fit results are distributed at the cadence of the data flow in the RTS (i.e., every \unit{8}{\second}).
Supplemental information for the polarisation conversion and regridding stages can be ``pre-computed'' based on array geometry and Earth rotation.
This is also disseminated to the worker nodes, but on slower time scales.

\begin{figure}
\centering
\includegraphics[height=7cm]{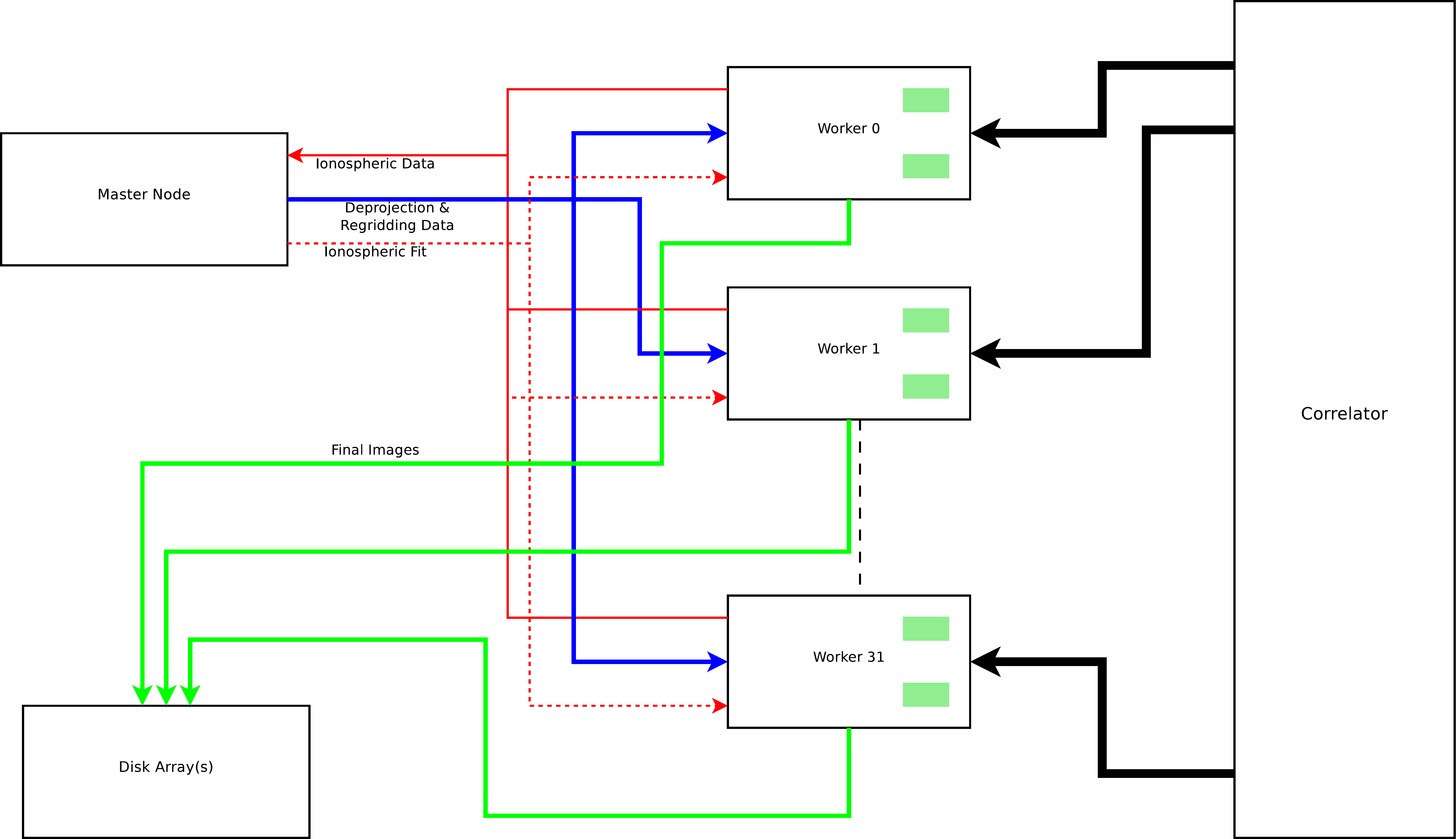}
\caption{Architecture of the Real Time Computer, which will run the Real Time System for the Murchison Widefield Array.
Each worker node contains two GPUs, and is connected to the correlator by two ethernet cables.}
\label{fig:RTCarch}
\end{figure}

Sixty four high end GPUs will consume approximately \unit{16}{\kilo\watt} of electricity.
If each host node consumes a further \unit{300}{\watt}, the total power consumption of the computational hardware will be approximately \unit{25.6}{\kilo\watt}, leaving spare capacity in the \unit{40}{\kilo\watt} power budget for the networking switches, disk arrays and head node.

Other hardware possibilities for the pipeline were considered, but rejected in favour of a system based on off-the-shelf GPUs.
We made a detailed study of MIPS processors, but found that the \FLOP{} requirements were too high, given the power constraints (as we shall see in Section~\ref{sec:singlegpu}, not even high end server processors come close to attaining the \unit{8}{\second} cadence).
A complete FPGA solution would almost certainly have a lower power consumption than the GPU solution, but was not practical.
The best example of this is the FPGA correlator constructed for the MWA.
Although the computational steps required for correlation are very well defined and understood, the scale of the MWA made construction of the correlator extremely difficult.
The tasks performed by the RTS are far less well defined (indeed, simply creating a single-pass algorithm for the calibration of the visibility data was a major scientific feat for the MWA), so the creation of an FPGA solution was deemed utterly infeasible.
A GPU solution allows far more flexibility in system design, and performance is certain to increase substantially on a relatively short timescale.
Similar considerations eliminated the Cell processor from consideration.
Although not as difficult to program as FPGAs, the Cell programming environment is still not as straightforward as GPUs.
Together with the cost advantages afforded by their origin as consumer hardware, a system based on GPUs was the option with the greatest probability of success.

\section{Approach to GPU Acceleration}
\label{sec:cuda}

The RTS is implemented in C, making the NVIDIA CUDA environment\cite{CUDASupercomputing07Tutorial,CUDAarticle1,GPUGEMS3} a natural choice for the GPU implementation\cite{2009arXiv0902.0915O}.
A prior feasibility study\cite{2007AAS...211.1104W} had demonstrated that GPU acceleration was beneficial for elements of the RTS, and this was used to guide the implementation of GPU processing for the entire pipeline.
Each pipeline stage was well contained, with defined inputs and outputs, enabling an incremental approach.
Inclusion of GPU acceleration is a compile-time decision, through the use of a pre-processor macro.
Continued maintenance of the CPU code is required to enable others without CUDA-capable GPUs or CUDA programming experience to contribute new algorithms to the RTS.

Within each pipeline stage, we identified large loops (typically over visibilities or pixels).
The loop bodies became CUDA kernels, and the loops themselves grids of thread blocks.
Such an approach requires that individual loop iterations be independent, which is generally the case throughout the RTS.
GPU data is held in `mirror' entries inside the relevant data structures.
Figure~\ref{fig:sampleimagestructure} shows how the data structure for an image contains a \texttt{data} field for the CPU data, and a \texttt{d\_data} field for the GPU data.
The allocate and release routines for each data structure were modified to manipulate GPU memory as well.

\begin{figure}
\begin{verbatim}
typedef struct _image_t {
  int sizex;
  int sizey;
  // Other fields [..]

  // Pointer to CPU data
  float *data[MAX_POLS];

#ifdef MWA_GPU
  // Pointer to GPU data
  float *d_data;
#endif
} image_t;
\end{verbatim}
\caption{A sample data structure for the CUDA-enabled Real Time System. Inclusion of the GPU data pointer is controlled by a preprocessor macro.}
\label{fig:sampleimagestructure}
\end{figure}

Extra routines transferred data between the GPU and CPU.
The cost of these transfers made it essential to implement the entire data processing pipeline on the GPU.
Often the actual transfer was relatively quick - the time was dominated by the need to pack data into pinned memory\footnote{Pinned memory is memory which cannot be paged to disk by the virtual memory system. It is essential for Direct Memory Access (DMA) transfers by the GPU} on the host machine (more on this below).
Transferring the data back and forth between the CPU and GPU at multiple points within the pipeline would negate the speed of the GPU.
This is particularly true for the CML, which consists of a number of smaller routines.

Even with the full pipeline implemented on the GPU, the transfers at the beginning and end of each \unit{8}{\second} cadence were still expensive, and extra efforts were made to minimise these.
The datastructure describing a collection of visibilities was a regular block of complex floating point values.
However, this was accessed with \verb+vis[i][j][k]+ pointer-to-pointer-to-array semantics, rather than the \verb|vis[k+ny*(j+i*nx)]| approach of computed indices.
The GPU obviously used the latter notation for performance, but converting between the two representations was slow.
The data first had to be copied into contiguous memory on the host, and then transferred to the GPU.
Changing the semantics of the entire CPU code was not practical given the time constraints placed on this project, so we made use of the \verb+amalloc+ library\citet{amalloc} to preserve the semantics in the majority of the code while allocating contiguous storage.
We then modified \verb+amalloc+ itself to use pinned memory, enabling faster DMA transfers to the GPU.

There were also portions of the code where  CPU code made use of an array of structures.
For the GPU, this had to be converted to a structure of arrays.
Initially, we performed the conversions on the CPU, and transferred the resulting arrays to the GPU.
However, this did not give satisfactory performance, so we moved the conversion to the GPU.
The GPU's increased memory bandwidth cut the time required dramatically.
Further gains would be possible through using pinned memory on the host, but this is not possible in the current CPU code.

This framework enabled rapid implementation of a GPU accelerated data pipeline.
Converting loop bodies to CUDA kernels was generally straightforward.
However, there were some exceptions to this, which had to be given special treatment.
The most significant of these was the gridder, which we will describe in detail in Section~\ref{sec:gridder}.

\section{Acceleration of the Gridder}
\label{sec:gridder}

The job of the gridder is to take the visibilities scattered across the $u$--$v$ plane, and interpolate them onto a regular grid prior to the FFT performed by the imager.
Mathematically, this is achieved by convolving each visibility, $V_{i}$ located at $(u_{i},v_{i})$ with a compact gridding convolution function (GCF):
\begin{equation}
V_{\text{grid}}(u,v) = \sum_{i} f_{i}(r) *  V_{i} \delta(u-u_{i}, v-v_{i})
\end{equation}
where the GCF $f_{i}(r)$ is zero for some $r > r_{\text{max}}$.
In the final system, $r_{\text{max}}$ is expected to be around twelve pixels, in order to enable correction of all of the structure in the telescope beam.
In current version of the RTS, we use a spheroidal function from \cite{1984iimp.conf..333S} as the GCF, which has the property of being its own Fourier transform.
In this case, all the $f_{i}$ are identical.
Ultimately, the optimal GCF will be determined by the CML, with a unique GCF for each visibility.

Programmatically, we compute the distance from the visibility to the centre of a given pixel, and then add the value of $f_{i}(r) V_{i}$ to that grid cell.
There are two ways to implement this procedure - as a scatter or a gather.
In a scatter operation, each visibility adds its contribution to the pixels within range, whereas in a gather, each pixel will find the visibilities in range and take a contribution from them.

The C code took the scatter approach.
For each visibility, the RTS calculated which $2 r_ {\text{max}} \times 2 r_ {\text{max}}$ patch of pixels it affected, and then added the weighted values to those pixels.
In the case of $r_{\text{max}} = 12$, each pixel is affected by around 60 visibilities.

However, scatter algorithms are not thread safe.
In the context of the gridder, the simple porting method described in Section~\ref{sec:cuda} fails dramatically.
If we give each visibility to a CUDA thread, a race condition develops whenever the  $2 r_ {\text{max}} \times 2 r_ {\text{max}}$ pixel patches associated with a pair of visibilities overlap;
we don't know which thread will write which value.
The CUDA API lacks atomic floating point instructions (although they are available for integers, and will be available for floating point variables in the upcoming ``Fermi'' GPUs), so this race condition results in undefined behaviour.
We verified this with a `scatter' kernel written in the manner outlined in Section~\ref{sec:cuda}.
The gridded data were dramatically different from the reference CPU implementation, with many visibilities prominently missing.
This is a problem which has been considered in connection with other parallel architectures \cite{CellGridder}.


\subsection{OpenGL Gridder}

While most of the hardware in NVIDIA GPUs used by graphics applications is also accessible via CUDA, there is some functionality that is currently available only via graphics APIs such as DirectX and OpenGL.
In particular, atomic floating point addition can be performed by rendering to the framebuffer with additive blending enabled.
With this ability, the gridder can be written as a scatter operation.
We experimented with an OpenGL-based gridder, which took advantages of specific features of the OpenGL graphics pipeline.
For each visibility, we simply rendered a screen-aligned quadrilateral texture-mapped with the appropriate GCF, under an orthographic projection and with additive blending enabled.
In addition to atomic accumulation in the framebuffer, this implementation takes advantage of fixed-function hardware for rasterization.
In this setting, the rasterizer efficiently computes which pixels are affected by a given visibility and spawns threads only for visibility-pixel pairs that contribute to the final result.
Computationally, the algorithm is $O(vk)$, for $v$ visibilities and $k$ pixels in each GCF.
Although this approach works well, those using, maintaining and developing future generations of the RTS will be experienced in C, but not OpenGL.
Accordingly, we developed an alternative approach in CUDA.


\subsection{CUDA Gridder}

The simplest gather algorithm for a CUDA gridder would give each grid pixel to a thread.
That thread would then examine each of the 130,000 visibilities in turn.
If a visibility were within range, its contribution would be added to the pixel.
We tested this approach and found that it produced correct results but took approximately \unit{20}{\second}.
This is not surprising given the calculation above; each pixel requires values from 60 visibilities (on average), but had to examine 130,000 to find them.
Each visibility incurred a full DRAM access latency (since GPUs lack caches), resulting in the extremely slow performance.

To reduce the search space, we binned the data.
We divided the pixel grid up into $24\times24$ pixel bins, and sorted the visibilities according to the bin in which they fell.
The bins were this size in  order to ensure that the largest GCF we were likely to encounter would be covered.
From the sorted list of visibilities, we built look up tables of the `first' and `last' visibilities within each bin.
When the CUDA gridding kernel started, each thread would compute the bin in which its own pixel fell.
The thread would then know that its pixel was only affected by visibilities in its own bin and the eight bins surrounding it.
The look up tables of `first' and `last' visibilities then enabled each thread to restrict itself to nearby visibilties, leading to substantial speed gains.
We further optimised the gridding by noting that patches of adjacent pixels (calculated by a single CUDA thread block) would lie within the same bin, and hence could collaborate.
Many visibilities could be loaded at once into shared memory, and used by all of the threads in the block.
We made use of the \texttt{thrust} library \cite{HoBe2009} to perform the visibility sort on the GPU, as well as for some other ancilliary operations.

The binning stage has complexity $O(v \log v)$ for $v$ visibilities, while the convolution stage has complexity $O(p)$ for an image size of $p$ pixels.
The prefactors for these scalings is controlled by the size of the gridding convolution function.
We are presently working on further optimising the gridder, to enable the binning to reflect the non-uniform distribution of visibilities in the $u$--$v$ plane.
This will be presented in a future paper.

\section{Benchmarks}
\label{sec:singlegpu}

We present a performance comparison between the CPU and GPU implementations of the RTS.
The CPU for all cases was a single core of an Intel Core-i7 920 (\unit{2.67}{\giga\hertz} ``Nehalem'') with \unit{6}{\gibi\byte} of memory.
For the GPU times, an NVIDIA Tesla C1060 was used.
The `master' process ran on a separate core of the same machine.

The worker processed twelve channels on its single GPU, following the design of Section~\ref{sec:RTCdesc}.
Fifty calibration sources were used by the CML in order to measure the tile gains and ionospheric distortion.
The gridder sorted the visibilities into $24\times24$ pixel bins, which is the effective gridding kernel size (note that this is larger than the gridding kernel used to produce Figure~\ref{fig:RTSflow}, but represents the largest convolution kernels we expect to require).
With a \unit{21.4}{\degree} field of view, $1125\times1125$ pixel final images were produced (matching the image size assumed in Figure~\ref{fig:RTSflow}).

In Table~\ref{tbl:rtsworktimes}, we show sample benchmarks for the computationally intensive portions of the code.
These timings are for particular computationally intensive portions of the code (the `work' time), and neglect all other operations (such as memory allocation and data transfer).
On the GPU side, these generally represent the time the GPU spends executing kernels.
Although most of the calculations are performed in single precision, the CPU code does make use of some double precision library calls.
When we reimplemented these libraries on the GPU, we converted them to single precision.
The calibration loop benefits enormously from GPU acclerations, and substantial gains are also made by the polarisation conversion and regridding stages.
The gridder has also become substantially faster, despite the drastic change in algorithm.

\begin{table}
\centering
\begin{tabular}{l|ccr}
Stage & CPU (ms) & GPU (ms) & \multicolumn{1}{c}{Speed up} \\
\hline
Main Calibrator Loop                      & 424790 &  2858  & $148\times$ \\ 
Gridding Convolution (CUDA)               & 13285 & 590  & $22\times$ \\ 
Imager FFT                                & 1564  &  172  & $9\times$ \\ 
Polarisation Conversion Engine            & 1572   & 46   & $34\times$ \\ 
Regridding Engine                         & 4104  & 147  & $28\times$ 
\end{tabular}
\caption{Comparison of the `work' times of various portions of the code. The `work' time of a pipeline stage is the main computational load with activities such as memory allocation and data transfer omitted.}
\label{tbl:rtsworktimes}
\end{table}

Table~\ref{tbl:rtstimings} shows timing information for the full code running with the parameters described above.
The timings include several extra steps which were not included in the timings of Table~\ref{tbl:rtsworktimes}.
For example, the `Imaging' line of Table~\ref{tbl:rtstimings} includes the `Imager FFT' line of Table~\ref{tbl:rtsworktimes}, plus allocation of temporary memory and the data re-ordering required by the two FFT libraries.
The Polarisation Conversion and Regridding times do not include the Precomputation phase for each of these stages, since these are not performed on every \unit{8}{\second} cadence.
The Calibrator Measurement Loop time includes the MPI communication with the master node, but this is irrelevant, due to the very small quantity of data sent and received (see Table~\ref{tbl:datasize}).
The difference in CML times between Table~\ref{tbl:rtsworktimes} and Table~\ref{tbl:rtstimings} is due to the initialisation (e.g. memory allocations and array zeroing) performed each \unit{8}{\second} cadence.

\begin{table}
\centering
\begin{tabular}{l|D{.}{.}{2}D{.}{.}{2}}
Stage & \multicolumn{1}{c}{CPU (sec)} & \multicolumn{1}{c}{GPU (sec)} \\
\hline
Acquire Data & 1.09 & 1.08 \\
Send GPU     & 0.0 & 0.03 \\
Calibrator Measurement Loop & 500.52 & 3.58 \\
Gridding Preparation & 5.13  & 0.17 \\
Gridding     & 14.70 & 1.40 \\
Imaging      & 3.78 & 0.34 \\
Receive GPU  & 0.0    & 0.05   \\
Polarisation Conversion & 3.56 & 0.10 \\
Regridding   & 4.55 & 0.49 \\
Cleanup      & 0.01 &  0.13 \\
\hline
Total & 533.34 & 7.37
\end{tabular}
\caption{Comparison of CPU and GPU timings for individual stages of the RTS. Timings are for a 12 channels, with the CML using 50 calibration sources. The gridding convolution function was $24\times24$ pixels in size, and $1125\times1125$ pixel images were produced. These timings do not include the precomputations for the Polarisation Conversion and Regridding stages.}
\label{tbl:rtstimings}
\end{table}

The porting process described here emphasised completeness over speed.
This was driven by the expense of memory transfers between host and GPU, as outlined in Section~\ref{sec:cuda}.
We have begun optimising the pipeline, in order to decrease the processing time further.
Some of the optimisations are related purely to the CPU side, some involve overlapping GPU and CPU execution, while other concern the GPU kernels themselves.

Although not directly related to the GPUs, data acquisition is a significant task in the RTS.
The `Acquire Data' item in Table~\ref{tbl:rtstimings} represents the time required to read the data from the system hard drive (or accept the correlator packets in the full system), and run the visibility integrator.
This is handled by a separate thread, so the reported time is how long the main pipeline thread is obliged to wait for its data to be ready.
It is evident that the data acquisition thread is having difficulty maintaining an adequate data rate in the current version of the code, since the pipeline thread has to wait over a second on each cadence.
If this time could be removed, larger images could be made; we estimate a \unit{30}{\degree} field of view could be set, producing $1600\times1600$ images.

The simplest optimisation to the GPU code is to ensure that all `workspace' arrays are pre-allocated at initialisation, and not allocated and released within each routine (this work has already begun, and is associated with the `Cleanup' line in Table~\ref{tbl:rtstimings}).
Storing 12 channels of data only requires around \unit{500}{\mebi\byte} of RAM, so there will be plenty of space remaining on a Tesla card for persistent `workspace' allocations.
The possibility of using the asynchronous CUDA API is also under consideration.
Currently, all kernel invocations are forced to be synchronous, but this is not really necessary given that each channel is almost entirely independent of all the others.
The processing of each channel typically proceeds as a set of small calculations to determine kernel arguments, followed by a kernel launch.
By using the asynchronous API, the calculation of the kernel arguments could be overlapped with the kernel invocation on the previous channel's data.
It would also be possible to stream each channel's data on to the GPU while processing started on the previous channel.

We have instrumented the CUDA pipeline itself, enabling us to measure the memory bandwidth usage and \FLOP{} count of each kernel.
We have found that very few kernels come close to the peak performance of the GPU in terms of both \FLOP{}s and memory bandwidth.
Over the entire pipeline, we have a sustained performance of about \unit{50}{\giga\FLOP\usk\reciprocal\second} and a sustained bandwidth (on the GPU itself) of about \unit{28}{\gibi\byte\usk\reciprocal\second}.
Note that when scaled to 64 GPUs, the implied performance is \unit{3.2}{\tera\FLOP\usk\reciprocal\second}, consistent with the \unit{2.5}{\tera\FLOP\usk\reciprocal\second} estimate of Section~\ref{sec:pipeline}.
We are currently working to increase the pipeline performance, in order to make it closer to the theoretical peak of the GPUs.
Some of the difference is certain to be intrinsic to the nature of the required computations, but we are confident that higher performance is possible.
For example, the order of array indexing on the GPU usually follows that on the CPU, but this often gives poor memory transaction coalescing on the GPU.
Re-ordering the arrays to maximise memory coalesing is certain to give substantial increases in performance.

The combination of all modules of the RTS, including supporting book-keeping operations must run end-to-end in under \unit{8}{\second}.
The GPU implementation described here achieves this goal for the runtime parameters described, while the CPU version does not (Table~\ref{tbl:rtstimings}).
Even processing a single channel on the CPU takes over forty seconds.
The overall speed up from using the GPU is around $72 \times$ with these parameters.
The GPU-accelerated pipeline is already in use at the MWA site, processing data from the prototype (consisting of 32 tiles).

\section{Conclusion}
\label{sec:conclude}

In this paper we have presented the results of our initial port of the MWA's data processing pipeline to NVIDIA GPUs.
This computationally intensive pipeline must operate on an \unit{8}{\second} cadence in order to keep up with the data flow from the instrument.
The original, single threaded, scalar CPU code took over \unit{500}{\second} to complete the processing of the data with reasonable operating parameters, and adding more CPUs to the system is not possible due to the electrical power constraints.
By implementing the entire data processing pipeline on the GPU, we were able to process all of the data required within the \unit{8}{\second} deadline.
Futher optimisation and future GPU hardware promise to increase both the size of images produced and the accuracy of the image calibration.
A GPU system is already on site, and processing data for the 32 tile prototype.

The Murchison Widefield Array is one of the next generation instruments to threaten a data deluge.
We have demonstrated how GPU acceleration can be used to manipulate and manage this flow.
As a technology demonstrator for the proposed Square Kilometer Array\cite{2006SPIE.6267E..76T}, whose processing needs are projected to be on the order of \unit{1}{\exa\FLOP\usk\reciprocal\second}, the MWA shows that highly parallel GPU processing is a practical part of a scalable heterogeneous computing platform for processing this data stream.


\section*{Acknowledgements}

We would like to thank NVIDIA for support via the Harvard CUDA Center of Excellence.
Funding support was provided in part by the Smithsonian Astrophysical Observatory.
This work was supported in part by the National Science Foundation through grants AST-0457585 and PHY-0835713.


\bibliographystyle{cpc}
\bibliography{mwa}


\end{document}